# The Revised Version of Class I Methanol Maser Catalog

I. E. Val'tts[1], G. M. Larionov[1], O. S. Bayandina[1,2]

## ABSTRACT

The revised version of the class I methanol maser catalog is presented. It contains 198 sources – new class I methanol masers detected in the direction of EGOs were added to the previous number (~160 sources have been published in the first version of this catalog – see reference in the text). Electronic version has been generated in the form of html file – http://www.asc.rssi.ru/MMI A statistical analysis was carried out within 2′ around a maser position to find an identification of class I methanol masers with any objects typical for star-forming regions – UCHII regions, IRAS sources, bipolar outflows, CS lines as of dense gas tracer, masers (class II methanol masers, OH and $H_2O$) and EGO. None of the bipolar outflow, already registered in the direction of class I methanol maser, did not coincide with EGO (with the exception of G45.47+0.07). The result is submitted in a form of a diagram.

## 1. Introduction

Interstellar methanol maser lines were accidentally discovered by Barrett et al. in 1971 using 37-m antenna in Haystack (USA) in the direction of well-known star-forming region Ori A. In the bandwidth of the $N_2O$ molecule, which they were looking for, Barrett et al. (1971) identified 5 strong lines of the ($J_2$-$J_1$) $E$-methanol series at a frequency of 25 GHz and expressed an assumption that the intensity of the observed lines have nonthermal nature. Then in the observations on the 100-m telescope in Effelsberg it was shown by Hills et al. (1975) that these narrow, bright lines are emitted by spatially separated components, the upper limit on the size of which gives the brightness temperature of more than 800 K, which exceeds by 10 times that maximum kinetic temperature, which can be obtained from the width of the methanol lines. The maser nature of the observed methanol lines was confirmed later in the interferometric experiment (Matsakis et al. 1980). At present we know 198 class I methanol masers and more than 500 class II methanol masers. It's about 700 possibilities, which should be used to study structure and kinematics of dust and gas in the interstellar medium and physical conditions around young stellar objects.

The classification of methanol masers, which was established by Batrla et al. (1987) and Menten (1987) was based on the next empirical fact: in the direction of some observed sources at some frequencies methanol maser lines are observed with full absence of emission (or rather, absorption lines or thermal emission) on the other frequencies, however in the directions of some other sources at the same frequencies - the opposite situation with of maser lines was observed. In fact, it was a manifestation of different pumping mechanisms: in some sources - collisional mechanism of inversion of the molecular levels operates (the I-st class), in others - collisional-radiative mechanism works (the II-nd class).

Interferometric observations have shown that class I methanol masers are isolated from the OH masers, $H_2O$ masers (up to 1 pc - Menten et al. 1986), from UCHII regions and infrared sources, while class II methanol masers are observed directly in the direction of UCHII regions and coincide at least, with the OH maser (see, for example, Reid et al. 1980). This property of the class I and class II methanol masers is the second fundamental feature of their differences.

---

[1] Astrospace Center of the Lebedev Physical Institue, Moscow 117997, Profsoyusnya 84/32.
[2] Moscow State Pedagogical University, Moscow 119991, M. Pirogovskaya, 1.

20 years ago, the pumping mechanism of class I methanol masers was understood as a simple consequence of the basic properties of methanol molecule itself. It was shown (Lees 1973), that in the result of collisional excitation of methanol an inversion is expected in the cascades of rotational levels of $J$ with the upper levels k = −1 in $E$-methanol and with the upper levels k = 0 in $A$-methanol. The preferred transitions would be k = −1-0 ($E$) and k = 0-1 ($A$) in accordance with selection rules at the frequencies 36 GHz ($4_{-1}$-$3_0 E$), 84 GHz ($5_{-1}$-$4_0 E$), 44 GHz ($7_0$-$6_1 A^+$), 95 GHz ($8_0$-$7_1 A^+$) and 146 GHz ($9_0$-$8_1 A^+$).

The complete similarity of spectra observed at these frequencies confirmed that these transitions are inverted by the same mechanism. The same mechanism generates the absorption lines at the frequency of 12.2 GHz ($2_0$-$3_{-1}$ E) (Batrla et al. 1987) and should form the absorption line at the frequency of 6.7 GHz ($5_1$-$6_0 A^+$) (Menten 1991a). Bright maser line, detected at 12.2 GHz (Batrla et al., 1987) and later, at 6.7 GHz (Menten 1991b), apparently are produced by another pumping mechanism pumping, and these masers belong to another class, which was named class II.

The Class I maser pumping mechanism does not require an additional energy source. However, as noted in the papers of Plambeck & Menten (1990) and Johnston et al. (1992), maser emission of this type may arise in the site of interaction of the bipolar outflow front with dense gas.

Radiative pumping of class II methanol masers was discussed in (Batrla et al. 1987), but a detailed collisional-radiative model was developed much later (see Cragg et al. 1992; Sobolev et al. 1994, 1997; and references therein).

Thus, the classification of methanol masers has following main points (Batrla et al. 1987; Menten 1991a,b).
The I-st class:
the emission in the transitions of $7_0$-$6_1 A^+$ at 44 GHz and $8_0$-$7_1 A^+$ at 95 GHz, and the absorption at the frequencies 12.2 GHz and 6.7 GHz, remoteness and isolation from UCHII regions, infrared sources, OH and H$_2$O masers, but possible association with bipolar outflows. Pumping mechanism is collisional. Prototype sources are the sources Ori KL, OMC2, NGC2264, W51, DR21West.
The II-nd class:
The emission in the transitions of $2_0$-$3_{-1}$ E (12.2 GHz), $2_1$-$3_0 E$ (19 GHz), $9_2$-$10_1 A^+$ (23 GHz) and $5_1$-$6_0 A^+$ (6.7 GHz), the association with UCHII regions, infrared sources, OH and H$_2$O masers, collisional-radiative pumping mechanism. Prototype sources are the sources W3 (OH), NGC7538, NGC6334E, F.

Although the first class I methanol masers were discovered in the direction of high-mass stars, it was suggested that just these masers, sufficiently remote from UCHII regions objects, and possibly associated with bipolar outflows, may be used to study the process of evolution of low-mass stars, in which the bipolar outflows play a dominant role. In contrast, class II methanol masers can be used to study hot and dense molecular cores in the vicinity of UCHII regions around massive stars.

In general, the established classification is correct so far, but now the situation is not so obvious. With the accumulation of observational data, it became clear that practically all main points of the classification above have exceptions. For example, in Walsh et al. (1997) and Slysh et al. (1999) it was shown that class I methanol masers correlate with UCHII regions very weakly, class I methanol masers and IRAS sources (Ellingsen et al. 1996) not associated at all (according to Szymczak & Kus (2000) - only 13% of cases). A correlation between the brightness of masers and IRAS sources is not observed (Van der Walt et al. 1996), although namely the radiation of UCHII regions and infrared sources must ensure that the pumping mechanism is radiative-collisional indeed.

No correlation was found between class I methanol masers and bipolar outflows (Kalenskii et al. (1992), while in some bipolar outflows, by contrast, class II masers were detected (Slysh et al. 1999). In addition, class I methanol was found at the frequency of 44 GHz in the direction of W3(OH) (Haschick et al. 1990), which is a classic example of class II and one of most powerful maser emitter at the frequency of 6.7 GHz (Menten 1991b). It is one of the prototype of class II maser which is based on the classification.

On the other hand, in surveys at 44 GHz (Morimoto et al. 1985; Haschick et al. 1990; Bachiller et al. 1990; Kalenskii et al. 1992; Slysh et al. 1994; Kurtz et al. 2004) and at 95 GHz (Ohishi et al. 1986; Val'tts et al. 1995; Val'tts et al. 2000; Ellingsen 2005) taken to trace methanol masers in the direction of class I masers a lot of class II methanol masers were found. More - in the interferometric studies with the VLA at 44 GHz Kurtz et al. (2004) showed that in the areas of massive stars, in which class II methanol masers are observed, class I maser emission at 44 GHz is also observed, with I and II classes coincide spatially within 0.2-0.5 pc. This was true even for the most powerful class II methanol maser G9.62+0.19 which previously was not supposed to manifest class I radiation.

However, statistically it was not checked and it is still unclear, whether deviations from the established classification are accidental or prevalent and systematic. To make such estimates, in 2007 we created a catalog of class I methanol masers (Val'tts & Larionov 2007), and we upgraded it now. The source sample and the description of class I methanol maser surveys were discussed earlier in Val'tts & Larionov (2007). Here we presented brief comments.

Already well-known class I methanol masers emit in the range from 9.9 GHz up to 229 GHz at 15 transitions of the rotational levels of methanol molecule. The most common and strong class I methanol masers are observed at the frequency 44 GHz in the transition $7_0$-$6_1A^+$ and at the frequency 95 GHz ($8_0$-$7_1A^+$). Surveys of star forming regions which have been carried out in order to search for class I methanol masers in other lines were less effective. Information about these surveys is, for example, in (Val'tts 1999). In the revised version we have compiled a complete catalog of class I methanol masers, based mainly on observations of the lines at 44 GHz $7_0$-$6_1A^+$ in the northern and southern hemispheres (Morimoto et al. 1985; Haschick et al. 1990; Bachiller et al. 1990; Kalenskii et al. 1992; Slysh et al. 1994; Kurtz et al. 2004). Class I masers that were detected at 95 GHz, but not observed at 44 GHz (Val'tts et al. 1995; Val'tts et al. 2000; Ellingsen 2005), five masers detected at 36 GHz (Liechti & Wilson 1996). 16 EGOs – 4.5 μm–selected outflows candidates (*Spitzer*, GLIMPSE, Cyganowski et al. 2008; Chen et al. 2009) - as new methanol masers detected at 44 GHz with the VLA (Cyganowski et al. 2009) – are also included. At the moment the revised version of the catalog contains 198 sources.

Twelve class I methanol masers were identified with GLIMPSE point sources detected as class II methanol masers in the survey made by Ellingsen (2007).

## 2. Catalog Description

A catalog is a table in the electronic form at http://www.asc.rssi.ru/MMI with a file 'readme' and a file of references. Search is for class II methanol masers observed in the direction of class I methanol masers and for class I methanol masers identified with EGOs.

The columns give:
(1) –    Original number of the class I methanol maser.
(2) –    Source position in Galactic coordinates.
(3) –    Sources name – optical or radio identification.
(4) & (5) – Equatorial coordinates for both epochs 1950/2000.
(6) –    For the brightest detail at 44 GHz - integrated line flux, Jy km/s
         (at least 3 Jy km/sec), & $V_{LSR}$, km/s, & reference.
(7) –    Maser identification: class II methanol maser (sign Y) OH and $H_2O$ masers.
(8) –    Infrared identification: IRAS number, IRDC (sign Y or N) – under construction, now

based only on the [E1, CWH, CES] - see references, EGO & GLIMPSE objects. (EGO 1 - "likely" MYSO outflow candidates, EGO 2 - "likely" MYSO outflow candidates in which confusion and/or saturation are problematic in some bands, EGO 3 - "possible" MYSO outflow candidates, EGO 4 - "possible" MYSO outflow candidates in which confusion and/or saturation are problematic in some bands, EGO 5 - outflow-only – abbreviation from Cyganowski et al. 2008),
GLIMPSE – GLIMPSE point sources detected as class II methanol masers in the survey made by Ellingsen (2007).

(9) –   Ultracompact HII region & bipolar outflow identification.
(10) –  CS(2-1) line identification.
(11) –  Distance, kpc & references.

Masers at 95 GHz marked with blue, masers at 36 GHz marked with red, wine color marked VLA observations.

### 3. Results of Statistical Analysis

A statistical analysis is presented in a form of a diagram (see Fig. 1). Sources marked in italics in the table (probable identification) are not included in our statistical estimations. Results: of 198 (100%) star-forming regions in which class I methanol masers have been detected
- in 79% of cases (157 sources) class I methanol masers are associated with IRAS sources (within the telescope beam and the mean IRAS positional accuracy);
- in 80% (158 sources) - with $H_2O$ masers;
- in 70 % of cases (138) sources class II masers are also observed;
- in 62% of cases (122 sources), class I methanol masers were identified with ultracompact HII regions;
- in 62% of cases (122 sources) class I methanol masers were identified with OH masers;
- in 60% of cases (118 sources) class I methanol masers were identified with CS line emission, which traces dense gas;
- in 30% of cases (60 sources) class I methanol masers were identified with EGO;
- in 20% of cases (40 sources) class I methanol masers were identified with bipolar outflows (nonregistrating EGO identification).

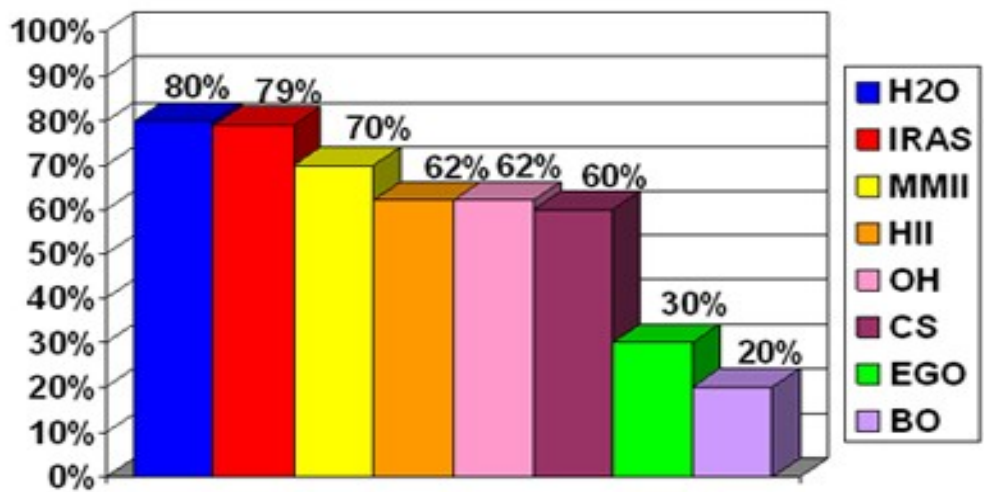

Fig. 1. Statistical analysis of class I methanol masers identifications.

## 4. Conclusions

- A new version of the class I methanol maser catalog was compiled and presented on the Internet: http://www.asc.rssi.ru/MMI.
- 198 class I methanol maser were catalogued.
- New 44 GHz masers found in recent survey in the direction of a new class of objects EGO (extended green objects) were added.
- Many methanol maser sources are objects of mixed type, combining classification features of both classes.
- In the revised version of the catalog more than 50% of class I methanol masers are associated with bypolar outflows – if outflows traced by EGO.
- None of the bipolar outflow, already registered in the direction of class I methanol maser, did not coincide with EGO (with the exception of G45.47+0.07).

This research has made use of NASA's Astrophysics Data System Bibliographic Services and the SIMBAD database operated at CDS, Strasbourg, France. Support for this work was provided by Russian Foundation of Basic Research (10-02-00147-a) and Program for Basic Research Division of Physical Sciences of the Russian Academy of Sciences "Active Processes in the Universe-2010".